\documentclass[groupedaddress,pra,showpacs,nofootinbib,superscriptaddress,longbibliography, notitlepage, twocolumn]{revtex4-1}

\usepackage{amsmath}
\usepackage{amssymb,bm}
\usepackage{amsthm,color,xcolor,dsfont}
\usepackage{graphicx} 
\usepackage[colorlinks=true, linkcolor=blue, urlcolor=blue, citecolor=blue, hyperindex, breaklinks]{hyperref}\usepackage{subfigure}
\usepackage{mathtools}
\usepackage{ulem}
\usepackage{float}
\newcommand\ieeeedit[1]{\textcolor{black}{#1}}

\newtheorem*{definition*}{Definition}

\def\equationautorefname~#1\null{Equation~(#1)\null}%

\newcommand{\ket}[1]{|#1\rangle} 

\begin{document}

\title{Fault tolerant resource estimation of quantum random-access
  memories}

\author{Olivia Di Matteo} \email[Electronic address:
]{odimatteo@triumf.ca} \affiliation{TRIUMF, Vancouver, BC, V6T 2A3,
  Canada} \affiliation{Department of Physics \& Astronomy, University
  of Waterloo, Waterloo, ON, N2L 3G1, Canada} \affiliation{Institute
  for Quantum Computing, University of Waterloo, Waterloo, ON, N2L
  3G1, Canada}

\author{Vlad Gheorghiu} \email[Electronic address:
]{vlad.gheorghiu@uwaterloo.ca} \affiliation{Institute for Quantum
  Computing, University of Waterloo, Waterloo, ON, N2L 3G1, Canada}
\affiliation{Department of Combinatorics \& Optimization, University
  of Waterloo, Waterloo, ON, N2L 3G1, Canada}

\author{Michele Mosca} \email[Electronic address:
]{michele.mosca@uwaterloo.ca} \affiliation{Institute for Quantum
  Computing, University of Waterloo, Waterloo, ON, N2L 3G1, Canada}
\affiliation{Department of Combinatorics \& Optimization, University
  of Waterloo, Waterloo, ON, N2L 3G1, Canada} \affiliation{Perimeter
  Institute for Theoretical Physics, Waterloo, ON, N2L 6B9, Canada}
\affiliation{Canadian Institute for Advanced Research, Toronto, ON,
  M5G 1Z8, Canada}

\date{Version of \today}

\begin{abstract}
  Quantum random-access look-up of a string of classical bits is a
  necessary ingredient in several important quantum algorithms. In
  some cases, the cost of such quantum random-access memory (qRAM) is
  the limiting factor in the implementation of the algorithm. In this
  paper we study the cost of fault-tolerantly implementing a
  qRAM. We construct and analyze generic families of circuits
    that function as a qRAM, discuss opportunities for qubit-time
    tradeoffs, and estimate their resource costs when embedded in a
    surface code.
\end{abstract}

\maketitle
\section{Introduction}

Random-access memory (RAM) is an essential component of classical
computing architectures. Many quantum algorithms require an analogous
system, a so-called quantum RAM, or qRAM, where the input is a quantum
state, the routing components are inherently quantum, and the
information stored can be either classical, i.e. $\ket{0}$ or
$\ket{1}$, or quantum, i.e. any arbitrary superposition of $\ket{0}$
and $\ket{1}$.  In the present paper we consider qRAM that stores only
classical information.  Generically, such a memory allows for the
querying of a superposition of addresses
\begin{equation}\label{eqn:qRAM}
  \sum_j\alpha_j\ket{j}\ket{0} \stackrel{qRAM}{\longrightarrow} \sum_j\alpha_j\ket{j}\ket{b_j},
\end{equation}
where $\sum_j\alpha_j\ket{j}$ is a superposition of queried addresses
and $\ket{b_j}$ represents the content of the $j$-th memory
location. A memory that stores classical information but allows
queries in superposition is required for quantum algorithms such as
Grover's search on a classical database \cite{PhysRevLett.79.325},
collision finding \cite{Brassard:1997:QCH:261342.261346}, element
distinctness \cite{Ambainis:2007:QWA:1328722.1328730}, dihedral hidden
subgroup problem \cite{Kuperberg:2005:SQA:1085579.1093661}, phase
estimation for electronic structure simulation \cite{Babbush2018}, and
various practical applications mentioned in
\cite{arunachalam2014quantum}. In fact, such a quantum memory often
plays the role of the oracle and is ideal in implementing any
oracle-based quantum algorithm, in which the oracle is used to query
classical data in superposition.

Several authors described algorithms that require only a polynomial
amount of resources such as computational qubits or depth, and some
larger number of `quantumly accessible' classical bits
\cite{10.1007/978-3-642-38616-9_6}. Such quantumly accessible
classical bits are less costly for classical simulations of quantum
computers, since a quantum algorithm with $n$ qubits and $m$ qRAM bits
can be simulated with $O(2^n + m)$ classical bits, instead of
$O(2^{n+m})$.

There are also numerous algorithms, for example in quantum machine
learning and Hamiltonian simulation, that are shown to demonstrate a
speedup but must assume a qRAM can be queried efficiently (for
example, \cite{Kerenidis2016, Wang2018, Zhao2018}). For many
algorithms this dependence stems from using the HHL algorithm for
solving linear systems as a subroutine \cite{HHL}, which itself is
successful in practice assuming we can query efficiently, though we
note that the particular operation performed by the qRAM differs
slightly from the one we analyze here \footnote{In contrast to
  \autoref{eqn:qRAM}, the data to be read in is a vector of complex
  numbers $\mathbf{b} = \left( b_1, b_2, ..., b_n \right)$ which
  become the amplitudes in a superposition, e.g.
  $\mathbf{b} \rightarrow \sum_i b_i |i \rangle$.} .

One qRAM implementation, the bucket brigade method
\cite{Giovannetti2008}, may allow algorithms that make only a few
queries to a qRAM to avoid the usual overhead associated with
fault-tolerant implementation of a binary-tree type look-up
circuit. For such few-query algorithms this may bring substantial
savings. However for many-query algorithms, it does not appear that
one can bypass fault-tolerant error correction
\cite{Arunachalam2015}. It remains unclear whether there is in fact
much of a savings over general purpose quantum memory when
implementing a fault-tolerant qRAM for a quantum computation.

Our work here seeks to address the question of the cost of
fault-tolerant qRAM in a quantum computation. This cost is comprised
of a number of different factors. The primary one we consider here is
the execution of a quantum circuit which performs the query,
completely embedded in a fault-tolerant error correction scheme. One
must also take into account external factors that may stem from the
specific physical implementation and/or algorithm which is querying
the qRAM. In principle, we can ``boot up" our qRAM qubits only when we
need to make a query, and turn them off after the fact. However, a
combination of the overhead cost of initializing the qubits, cost of
resetting them to $\ket{0}$, as well as any idle time between queries
may warrant a more active approach to error correction so that after
initialization the qRAM remains perpetually on (in a sense this would
be similar to how conventional RAMs refresh themselves).

Roughly speaking, there are two natural ways to implement a quantum
query to a classical memory. At one extreme, classical information
$b_1, \ldots ,b_N$ could be \ieeeedit{explicitly} laid out in static physical hardware which
is quantumly queried. This can be accomplished using, for example,
controlled-NOT (CNOT) gates on some target register conditioned on the
values of the $b_j$, or by some binary tree circuit or a bucket
brigade-style circuit, which has depth logarithmic in the size of the
database. Such approaches require a number of query qubits that is
proportional to the size of the database. By query qubits, we are
referring to the qubits used in essentially all such memory schemes in
order to connect the computational qubits that store the index that
needs to be queried, and the (qu)bits that store the classical
information being queried. These query qubits (or qudits) are only
used ephemerally in order to perform the query, and do not store any
information before or after each query.

\ieeeedit{Such an `explicit' qRAM has the advantage that a circuit implementing it needs to be compiled and optimized only once, while the contents of the memory are free to change. The disadvantage is the significant space overhead required, since we need as many qubits as we have bits in the database, and these qubits need to be initialized and maintain coherence.}

At the other extreme, instead of storing the classical information in
a static physical memory, one can simply implement a sequence of
mixed-polarity multi-controlled CNOTs conditioned on the control bits
representing the memory address of a 1 \footnote{Of course, if there
  are more 1s than 0s, it suffices to condition instead on the control
  bits representing the memory addresses of the 0s and then finish
  with a single NOT gate. Later in the paper we discuss other
  optimizations that can be performed based on structure or patterns
  in the data.}. This implies that the classical database is
implicitly stored in the logical circuit, and this circuit will have
depth proportional to the number of 1s in the database. 

\ieeeedit{An advantage of `implicit'
circuits is that they can be heavily optimized using any number of known
optimization techniques for Boolean circuits. However, an implicit qRAM requires us to know the contents of the memory in advance. Writing to the qRAM consists of the addition or removal of a multi-controlled CNOT corresponding to the desired address, meaning that any change in the database would require recompilation and optimization of the circuit.}

We present variants of this latter approach that only require
resources roughly proportional to the number of 1s in the
database. Suppose there are $m$ 1s in the database. We can perform a
sequence of $m$ multiple controlled gates, where each address
$x_1 x_2 \ldots x_n$ containing a $1$ is used to control the output of
some target bit. We can also parallelize this process: in $O(\log m)$
depth we can compute $m$ copies of the desired index. Then in
parallel, using the $j$-th copy of the desired index we compute $1$ if
the index equals the index of the $j$th 1 in the database. Finally,
using a binary tree type circuit we can in $O(\log m)$ depth compute
whether a $1$ was computed on any of the $m$ copies. We then uncompute
all the intermediate computations.

There are also many natural ways to interpolate between the two
approaches, for example using the same fan-out like operation to make
$2^k$ copies of the first $k$ index bits, and then use $2^k$ parallel
logical circuits to explore the remaining $n-k$ index bits.

In this paper, we outline these various questions and approaches, and
consider their costs and trade-offs. Such an analysis is important for
optimizing the physical resources needed to implement in practice
quantum algorithms that use a qRAM repeatedly, using the best-known
methods. We begin by describing our general cost model and how we
presume an algorithm will query the qRAM. We then introduce a number
of circuit families: circuits for the bucket brigade qRAM model, the
basic highly sequential or parallel circuits mentioned above, as well
as some interpolations between the two. Under assumptions about our
physical parameters and our database, we will compute concrete
parameters such as real-time cost, number of qubits, etc. of our
circuits embedded within a fault tolerant implementation using a
defect-based surface code \cite{Fowler2012}. Following this, we apply our techniques
to a family of circuits recently designed by other researchers \cite{Vadym2018}, demonstrating 
how targeted optimization can significantly reduce the resources costs. We conclude with some
final thoughts and present avenues for future research. Finally, we
note that we have made available a code repository including our data
as well as circuit details and resource estimation procedures
\cite{code}.


\section{Modeling the cost of a qRAM}

We consider a qRAM that stores quantumly accessible classical
bits. Locations in memory are addressed by $n$-bit strings
$x_1 x_2 \cdots x_n$, and are queried by inputting the associated
state $\ket{x_1 x_2 \cdots x_n}$ to a circuit. The memory contents may
be stored explicitly in additional qubits which are coupled to during
the query, or implicitly in the circuit given prior knowledge of the
database. In both models, for each address $x_1x_2\cdots x_n$ the qRAM
should implement
\begin{equation}
  \ket{x_1 x_2 \cdots x_n} \ket{0} \rightarrow \ket{x_1 x_2 \cdots x_n} \ket{b_{x_1x_2\cdots x_n}} 
  \label{eqn:qRAM_mapping}
\end{equation}
\noindent where $b_{x_1x_2\cdots x_n}$ is the stored value at the
specified address. This form ensures the qRAM can be queried in
superposition.

\

\begin{figure}[h]
  \centering
  \includegraphics[width=1\columnwidth]{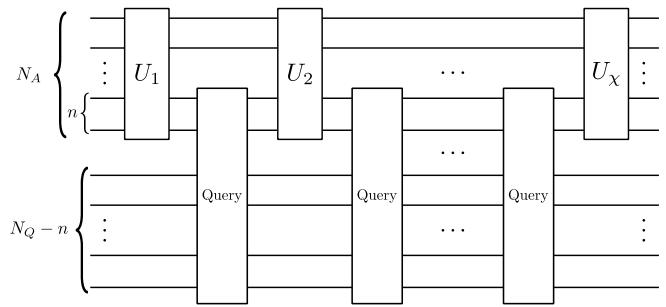}
  \caption[Schematic of algorithm querying a qRAM]{An example of how
    an algorithm might query a qRAM. The algorithm itself runs on
    $N_A$ logical qubits, $n$ of which will be used as an input
    address to query a qRAM, which itself requires $N_Q$ logical
    qubits. We suppose that the algorithm queries the qRAM
    periodically after performing each of some number $\chi$ of
    unitary operations.}
  \label{fgr:algorithm_schematic}
\end{figure}

Consider a quantum algorithm with a structure similar to that in
\autoref{fgr:algorithm_schematic}. Queries to a qRAM are interspersed
between some number of arbitrary unitary operations that comprise the
main portion of the algorithm. We suppose that the entire circuit is
embedded in a surface code in order to make it fault-tolerant. Then,
we can use the same cost metric as \cite{SHAPaper}, wherein
\begin{equation}
  \hbox{Cost} = \log_2 \left( \hbox{Logical qubits} \times \hbox{Surface code cycles.} \right)
\end{equation}
In essence this cost represents a tradeoff of space vs. time.

The cost is calculated using a framework \cite{SHAPaper} that starts
from the high-level algorithmic description. The algorithm begets a
quantum circuit, which is then synthesized and optimized over an
elementary gate set (we use Clifford+$T$). The optimized circuit is
then embedded into a surface code, in which the number of logical
qubits and operations determine parameters such as code distance and
resources required for magic state distillation. With these, we can
compute physical layer parameters such as the number of physical
qubits and surface code cycles.

In the present work we will take a more general approach to
calculating cost. \ieeeedit{We will compute the cost of an `isolated' qRAM
circuit making a large number of queries and express it in terms of
parameters such as $n$, and the number of 1s stored in the memory. We will not be dealing with any specific algorithms or circuits. As a consequence, we cannot
perform the kind of extensive optimization that may be afforded to us when we know the memory contents. We can, however, consider higher-level optimizations, such as the choice of decomposition of the mixed-polarity gates.}

We will also use a `rough' estimate of cost before performing the
surface code analysis:
\begin{equation}
  \hbox{Rough cost} = \log_2 \left( \hbox{Logical qubits} \times \hbox{$T$-depth} \right).
  \label{eq:rough_cost}
\end{equation}
$T$ gates are the most expensive part of the implementation due to the
need to distill magic states, and the algorithm will be time-limited
to how quickly we can produce these states. Magic states are distilled
in separate surface codes called distilleries; once a state has been
distilled, it can be held in reserve and injected into the circuit
when needed \cite{Fowler2012}. In the interest of minimizing time, it
is beneficial to have multiple distilleries running simultaneously,
especially since in many of our circuits we will need to run $T$ gates
in parallel. However, adding distilleries incurs an additional cost
due to the increased number of physical qubits. One must therefore
choose this quantity based on the available resources.

In our analysis we select the number of distilleries by considering
the average $T$-width $T_w = T_c / T_d$, where $T_c$ is the total
$T$-count of the algorithm, and $T_d$ is the $T$-depth, i.e. the
number of layers of depth that contain $T$ or $T^\dag$ gates. In many
of the circuits we present, we use repeated copies of sub-circuits
(e.g. a Toffoli) which have been pre-optimized such that they have
$T_d = 1$, or close to 1. Then the number of gates in each layer of
$T$-depth should be roughly the same, and we will make efficient use
of our factories as we will essentially be consuming the required
magic states immediately upon creation, using the minimal number of
physical qubits required to do so.

\section{Should I try turning it off and on again?}

One question posed in the Introduction was whether or not we should
keep the qRAM ``on'' between queries. By this, we mean performing
active error correction on the idle qubits in the qRAM while the
algorithmic components run. We will consider this question in terms of
the difference in cost of both approaches.

For simplicity, let us assume that the algorithm performs $\chi$
instances of the same operation $U$ on $N_A$ qubits, and $\chi - 1$
queries to the qRAM in between these operations, using $N_Q$ qubits,
as is depicted in \autoref{fgr:algorithm_schematic}. As we alternate
between queries and operations, the logical qubits in the algorithm
(minus the $n$ used as an address for the query) will be in an idle
state while we query the qRAM; similarly, if we keep the qRAM on, the
query qubits will be idle while the algorithmic portions run.

Let $c_{A}$ be the number of surface code cycles taken by one instance
of $U$. We assume that $c_i$ cycles are required for the
initialization of a logical qubit, using for example the procedures
outlined in \cite{Fowler2012}. We assume as well that it takes $c_t$
cycles to reset and dispose of the qubits once the query is
complete. In a sense, these operations are analogous to C memory
management functions \texttt{calloc()} and \texttt{free()}
\cite{KR:1988:C}.

It is straightforward, from \autoref{fgr:algorithm_schematic}, to see
that we should take the following approach:
\begin{eqnarray*}
  c_A &>& c_i + c_t  \rightarrow \hbox{turn off qRAM qubits while running $U$,} \\
  c_A &<& c_i + c_t  \rightarrow \hbox{always keep qubits on}.
\end{eqnarray*}
Note, however, that since cost involves multiplication by the number
of logical qubits, performing any sort of cost analysis on a qRAM is
naturally most important when $N_Q >> N_A$. The choice of query
circuit, various options for which will be discussed in the ensuing
sections, may thus depend on a tradeoff between available resources
and the relative size of the algorithm circuit versus that of the
qRAM.

One can also imagine situations in more complex algorithms where the
operations $U_i$ are not identical. Here one might design a smart
compiler that will choose to periodically turn off the qRAM during
comparatively long algorithm operations, but leave it on for shorter
ones. A smart compiler may also take into consideration the relative
error rates between, say, re-initializing a qubit in $|0 \rangle$
versus keeping its existing state error-corrected. This alone may
warrant always turning off the qRAM, if $| 0 \rangle$ can be
initialized quickly and with very high accuracy.

Finally, we note that if an algorithm requires writing regular updates
to the classical database (such as in
\cite{10.1007/978-3-642-38616-9_6}), then one would need to update the
query circuit during the course of the algorithm. Thus any latency
between changing the database based on a measurement during the
execution of the quantum circuit and updating the circuit in the
software must be considered. Latency would also exist for any other
mechanism for storing the database, and the precise cost for each
approach would need to be considered in each case.


\section{Bucket brigade circuits}
\label{sec:bucketbrigade}

\begin{figure}
  \centering \includegraphics[width=0.95\columnwidth]{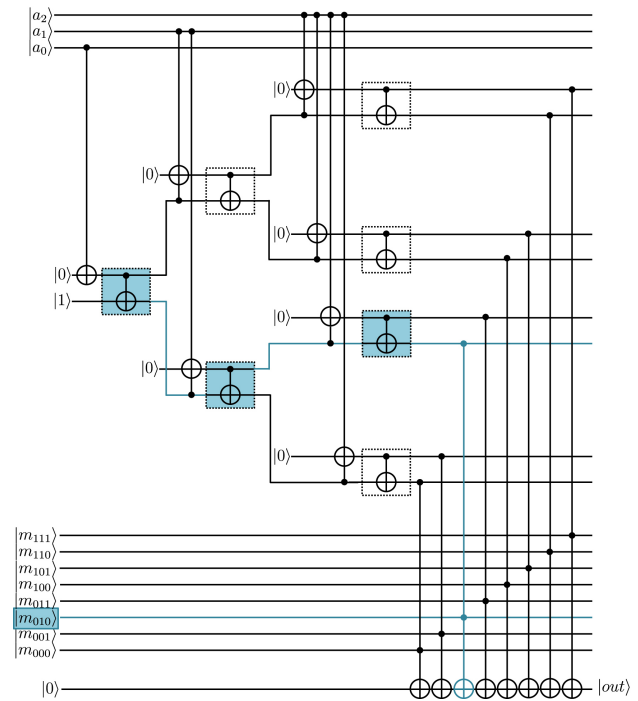}
  \caption[A bucket brigade qRAM circuit.]{One method of constructing
    a bucket brigade style qRAM circuit. Original image taken from
    \cite{Arunachalam2015}. The circuit is implemented using only
    CNOTs and Toffolis, which we decompose over Clifford+$T$. The
    circuit is independent of the contents of the memory, which is
    initialized separately in the lower register. For a fully
    reversible qRAM query we must run this circuit as written, and
    then uncompute the fanout of the address bits}
  \label{fig:bucketbrigade}
\end{figure}

The first family of circuits we will analyze are a family introduced
in \cite{Arunachalam2015} that \ieeeedit{implement} a bucket brigade qRAM
\cite{Giovannetti2008, Giovannetti2008b}. One \ieeeedit{such circuit} is shown in \autoref{fig:bucketbrigade}. These circuits
assume that the contents of the memory are stored statically in the
lower register of qubits in \autoref{fig:bucketbrigade}, in contrast
to the circuits we will see in later sections. They are thus not
perfectly comparable, however it is of interest nonetheless to
estimate the resources required when a bucket-brigade circuit is
implemented fault-tolerantly in order to deal with errors when making
a large number of queries.

Bucket brigade circuits are constructed using only CNOTs, and
Toffolis, which we will further decompose over the Clifford+$T$ gate
set. We will perform a cost estimate for two types of bucket brigade
circuit, those constructed exactly as in \autoref{fig:bucketbrigade},
as well as an improved version where we parallelize the execution of
the Toffolis in each layer.

\ieeeedit{We can choose from a number of
different implementations of a Toffoli \cite{Selinger2013,Amy2014},
which will affect the overall resource counts:}
\begin{itemize}
\item No ancillae, $T$-depth 3,
\item One ancilla, $T$-depth 2,
\item Four ancillae, $T$-depth 1.
\end{itemize}
Note that in all cases the ancillae can be reused whenever Toffolis
are applied sequentially. We focus on the third option, which while it
uses the most ancillae, a simple analysis using the cost function in
\autoref{eq:rough_cost} found this to always produce the lowest cost
due to the savings in $T$-depth.

\ieeeedit{To perform our resource estimates, we assume that the Toffolis 
in \autoref{fig:bucketbrigade} are all implemented using the decomposition in Fig. 1 of
\cite{Selinger2013}, which has $T$-count 7, $T$-depth 1, 16 CNOTs, and depth 7.
For a fully reversible qRAM query, we must run the circuit in
\autoref{fig:bucketbrigade}, and then uncompute the fanout of the
address bits. For memories with $n$-bit addresses, a bucket brigade circuit thus requires
$3\cdot 2^n - 4$ Toffoli gates. This yields resource counts:
\begin{eqnarray}
  N_Q &=& n + 2^{n+1} + 5 \quad \hbox{(logical qubits)}, \nonumber \\
  D &=& 21 \cdot 2^n + 2n - 26 \quad \hbox{(depth)},  \nonumber\\
  T_c &=& 21\cdot 2^n - 28 \quad \hbox{($T$-count)}, \nonumber \\
  T_d &=& 3\cdot 2^n - 4 \quad \hbox{($T$-depth)},  \\
  \hbox{CNOT}_c &=& 50 \cdot 2^n - 64 \quad \hbox{(CNOT count)}.  \nonumber
\end{eqnarray}}

We note here that this is only one possible implementation of the
bucket brigade scheme, and it is not necessarily the cheapest. If the
main algorithm querying the qRAM is concerned with phases (such as in
Grover's algorithm), it would be best to directly implement the
phase-shift version of the qRAM, i.e.
$\ket{j} \longrightarrow (-1)^{b_j} \ket{j}$. We could accomplish this
by replacing the final sequence of Toffolis with controlled-phase
gates from each address fanout qubit down to the corresponding memory
location. This would significantly reduce the circuit depth and
$T$-depth in the circuit of \autoref{fig:bucketbrigade}, and also
eliminate the need for the final output qubit. \footnote{Conversely,
  it is known how to convert phase-shift queries to bit-flip queries
  at the logical level as well, though with a small overhead. In a
  concrete application, the subtle cost differences between the two
  approaches should be evaluated to decide whether overall a
  phase-flip qRAM or bit-flip qRAM is more efficient. For this paper,
  we focus on the slightly harder bit-flip look-up.}

We naturally can also parallelize this circuit by a) copying down the
address qubits to ancillae so that the Toffolis in the address fanout
layers can be performed in parallel, and b) adding an extra output
register of $2^n$ qubits in the superposition of all even-parity
states to collect the results of the final set of Toffolis, followed
by CNOTs to compute the parity (as is depicted later in
\autoref{fgr:wide_shallow_evenparity}). We then copy down to an output
bit, apply another CNOT back to the even-superposition register, and
then uncompute the parity as well as the address bit fanout.

We need to prepare the even-parity superposition only once at the
beginning, and uncompute it at the end, as by design this register
will remain in superposition after each query. We thus neglect the
resources for this procedure in our analysis, as it can be performed
in logarithmic depth and becomes negligible for large $n$.

We require enough ancillae to perform the largest amount of
simultaneous Toffolis, which will be $4 \cdot 2^n$ using the $T$-depth
1 implementation. While in absolute terms this adds an exponential
number of qubits, it will yield a significant savings in $T$-depth
and, we will see, reduce the overall cost.  

\ieeeedit{For computing the circuit depth, note that computation of parity in the output register can be done at the same time as uncomputation of the address fanout portion, so it does not contribute to the depth. Furthermore, the initial copying of the address bits (so that the fanout Toffolis can be run in parallel) requires in the worst case $n - 1$ layers of depth, all of which can be absorbed in the calculation of the preceding Toffolis. Thus, the depth of the fanout portion is simply the depth of $n-1$ Toffolis, plus $n + 1$ for the initial CNOTs and layers in between the Toffolis.}

\ieeeedit{ Putting together the resource counts for the entire circuit, we obtain:
\begin{eqnarray}
  N_Q &=& 8 \cdot 2^n  \nonumber\\
  D &=& 16n - 5 \nonumber\\
  T_c &=& 21 \cdot 2^n - 28,  \nonumber\\
  T_d &=& 2n - 1, \\
  \hbox{CNOT}_c &=& 54 \cdot 2^n - 2n - 66. \nonumber
\end{eqnarray}}
The rough cost scaling of $N_Q \times T_d$ in terms of $n$ is greatly
improved in the parallelized circuit, and so moving forward we will
consider only the parallel version.


\section{Basic query circuits}
\label{sec:basic}

We now construct families of `implicit' qRAM circuits that will
highlight the space-time tradeoff required for a fault-tolerant
qRAM. We will see in the end that they have comparable overall cost,
but vastly differing use of resources down at the physical level.

For these circuits, we suppose for simplicity that the memory contains
$2^q$ 1s, the locations of which are known, with the rest being
0. Note that this is in contrast to the bucket brigade circuits where
the contents, while unknown, were provided to us statically stored in
hardware. Recall that if the number of 1s is ever greater than the
number of 0s, we can equally well build our circuits by inputting the
locations of the 0s.

We consider a small running example for the purpose of creating the
circuit diagrams. Suppose we have $n = 3$ and $q = 2$, i.e. 4 of 8
memory locations store a 1. We arbitrarily set those locations to be
000, 001, 011, 111.

\subsection{Large depth, small width circuit}

\begin{figure}[H]
  \centering \includegraphics{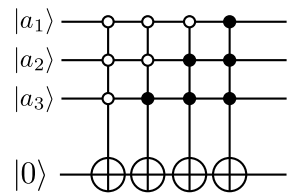}
  \caption[A qRAM circuit with few qubits but large depth.]{A qRAM
    circuit with few qubits but large depth. The addresses of all
    $2^q$ locations known to contain a 1 are implicitly stored in the
    circuit as mixed-polarity multiple control Toffoli gates.}
  \label{fgr:largedepth_dumb}
\end{figure}

We can easily create a circuit that outputs 1 for the valid addresses
by implementing a sequence of $2^q$ $n$-bit mixed-polarity multiple
control Toffolis (MPMCTs). The circuit for the running example is
shown in \autoref{fgr:largedepth_dumb}. Each MPMCT is tied to one of
the addresses, and sets a target bit to 1 only if its associated
address is fed in.  Henceforth we assume that we have a random
database where we don't know the exact sequence of MPMCTs, only that
we have $2^q$ such operations. As such, \emph{we will not perform any
  extensive circuit optimization in our analysis}, and will focus on
this worst case.

In general, when a sequence of MPMCTs \emph{is} known, we may be able
to greatly simplify the circuit. For example, since a sequence of
MPMCTs represents a Boolean function which is a sum of product terms,
we can find its ESOP expression using a tool such as EXORCISM-4
\cite{Mishchenko2001}. This can offer great savings - for example, the
circuit in our example can be reduced to 2 Toffolis just by factoring
some terms in the Boolean expression. We can also use the method of
\cite{Shafaei2013} which first computes the ESOP, and then breaks the
expression down into common cofactors of the expression terms;
cofactors are then reversibly synthesized individually before being
used in their constituent terms.

In our worst-case analysis, we begin instead by performing quantum
circuit synthesis to decompose the circuit down into the 1- and
2-qubit operations of the Clifford+$T$ gate set. Again, when the
precise sequence of MPMCTs is known, further optimization can be done
to reduce parameters such as the $T$-count and $T$-depth (see, for
example, the methods in \cite{Amy2014}).
 
The decomposition we choose for the MPMCTs is \ieeeedit{that in Fig. 4 of
\cite{Abdessaied2016a}}, which is an optimization of an
older algorithm from \cite{Barenco1995} that performs an
$n$-controlled NOT using $n - 2$ ancillae by turning it into a cascade
of $4(n-2)$ Toffoli gates. \ieeeedit{We take advantage of an additional
optimization which, at the cost of one more ancilla qubit, can further
parallelize some of the $T$ gates in the constituent controlled-phase gates (see (9) in \cite{Selinger2013}).} While it may be
possible to implement such gates with fewer ancilla qubits, recall
that we are interested in plugging this circuit into the surface code,
wherein minimization of $T$-count and $T$-depth plays a critical role
in reducing the overall cost incurred by magic state
distillation. Finally, we note that this MPMCT implementation is valid
only for $n \geq 4$, and so we will limit our analysis to this case.

\ieeeedit{Using the constructions of \cite{Abdessaied2016a} and \cite{Selinger2013}, we calculate the resources required for an $n$-controlled MPMCT gate:}
\begin{eqnarray}
  D &=& 28n - 60, \nonumber \\
  T_c &=& 12n - 20, \nonumber \\
  T_d &=& 4(n - 2), \\
  H_c &=& 4n - 6 \nonumber,\\
  \hbox{CNOT}_c &=& 24n - 40,  \nonumber
\end{eqnarray}
\ieeeedit{where $H_c$ is the number of Hadamard gates}.

We see immediately that in theory, the circuit in
\autoref{fgr:largedepth_dumb} requires only $n + 1$ qubits, plus $n-1$
ancilla qubits. As the ancillae are returned to their initial state
after each MPMCT, we can reuse them for all $2^q$ gates. We note that
any $X$ gates to change polarity of the controls can be applied in the
first layer of depth of each MPMCT; in addition, the final polarity
change can be performed in the last layer of the last MPMCT. In all
cases the $X$ gates do not contribute to depth, and as they are not
critical to the overall cost in a fault-tolerant setting, we will
disregard them. Thus we obtain total logical resource counts
\begin{eqnarray}
  N_Q &=& 2n,  \nonumber \\
  D &=& 2^q(28 n - 60),  \nonumber\\
  T_c &=& 2^q (12 n - 20), \nonumber \\
  T_d &=& 2^{q+2}(n - 2), \\
  H_c & = & 2^q (4n - 6), \nonumber\\
  \hbox{CNOT}_c &=& 2^q(24n - 40).  \nonumber
\end{eqnarray}

\subsection{Small depth, large width circuit}

\begin{figure}[h!]
  \centering
  \includegraphics[width=\columnwidth]{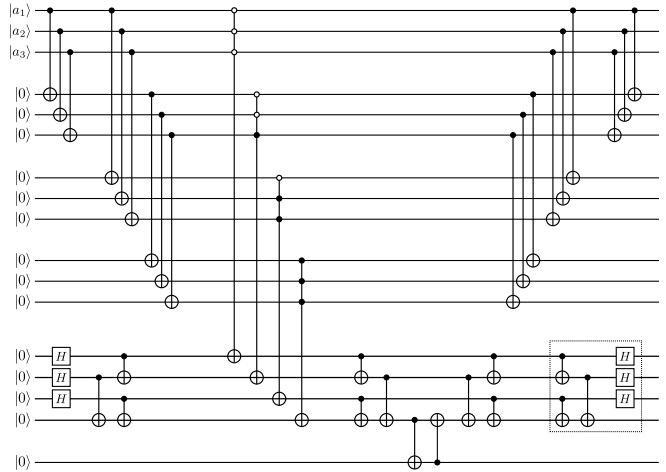}
  \caption[A qRAM circuit with small depth but many qubits.]{A circuit
    with small depth but a large number of qubits, where the
    implementation of the MPMCTs is performed in parallel. Each MPMCT
    requires $n - 1$ ancillae, significantly increasing the number of
    qubits. The register containing Hadamards prepares the
    superposition over all even-parity states, which eliminates the
    need to uncompute the MPMCTs when making the query fully
    reversible. When multiple queries are performed, we do not need to
    implement the portion of the circuit in the dotted box, as the
    uncomputation of the parity following the CNOT from the output bit
    leaves us in the even-parity superposition; the dotted box serves
    to return these qubits to $\ket{0}$. Resources for the creation
    and destruction of the even-parity superposition are neglected
    from our analysis, as this need only be performed once, and can
    then be used for a large number of queries.}
  \label{fgr:wide_shallow_evenparity}
\end{figure}

In contrast to the circuit of the previous section that performs all
the MPMCTs sequentially, here we present an implementation that
parallelizes their execution. The circuit for our running example is
shown in \autoref{fgr:wide_shallow_evenparity}.

We begin with $2^q$ registers of qubits, one for each address
containing a 1. The address is input to the first register of qubits
and then copied down to the others using a log-depth cascade of
CNOTs. Each register performs an MPMCT which triggers one of the
qubits in an additional register of $2^q$ qubits if the input address
matches. This register is prepared in a superposition over even-parity
states, as proposed in \autoref{sec:bucketbrigade}. We then compute
the parity, copy it to an additional qubit, copy back, and then
uncompute parity and the address fanout. We again neglect the
resources required to prepare the even-parity superposition, as this
need only be done once when performing multiple queries.

The number of qubits, including required ancillae, is
\begin{eqnarray}
  N_Q &=& n \cdot 2^q + 2^q(n - 1) + 2^q + 1 \notag\\
      &=& n 2^{q+1} + 1.
\end{eqnarray}

To compute the depth of this circuit, we must take into account the
sequences of CNOTs. The initial copying is performed in depth $q$, the
parity in depth $q$, plus 2 for copying to the output qubit. \ieeeedit{As uncomputation of the
address fanout has the same depth as parity, we can perform
them simultaneously}. Thus the
depth is
\begin{eqnarray}
  D &=& q + (28 n - 60) + q + 2 + q \notag \\
    &=& 28n + 3q - 58
\end{eqnarray}

\noindent which scales linearly in both $n$ and $q$.

The $T$-count will be the same as for the circuit in the previous
section, however the $T$-depth will now be $T_d = 4(n-2)$ as all the
MPMCTs are performed in parallel.

In addition, we will need the Clifford counts. The number of Hadamards
is unchanged (neglecting the construction of the even-parity
superposition). The number of CNOTs in the initial fanout plus a
parity computation is
\begin{eqnarray}
  \hbox{CNOT}_{c-init} &=& n \sum_{i=0}^{q-1} 2^i + \sum_{i=0}^{q-1}2^i \notag\\
                       &=& (n + 1)(2^q - 1) 
\end{eqnarray}

Doubling this, adding 2 for the copying to the output, plus the
$24n - 40$ CNOTs in the MPMCTs, we obtain
\begin{eqnarray}
  \hbox{CNOT}_c &=& 2(n + 1)(2^q-1) + 2^q (24n - 40) + 2\notag \\
                &=& 2^q (26n - 38) - 2n
\end{eqnarray}

In summary, the resources required are

\begin{eqnarray}
  N_Q &=& n 2^{q+1} + 1, \nonumber \\
  D &=& 28n + 3q - 58,  \nonumber \\
  T_c &=& 2^{q} (12n - 20), \\
  T_d &=& 4(n-2),  \nonumber \\
  H_c &=& 2^q (4n - 6),  \nonumber \\
  \hbox{CNOT}_c &=&  2^q (26n - 38) - 2n  \nonumber
\end{eqnarray}

\subsection{Preliminary cost estimate}
\label{subsec:preliminaryest}

\begin{figure}[h]
  \centering
  \subfigure{\includegraphics[width=\columnwidth]{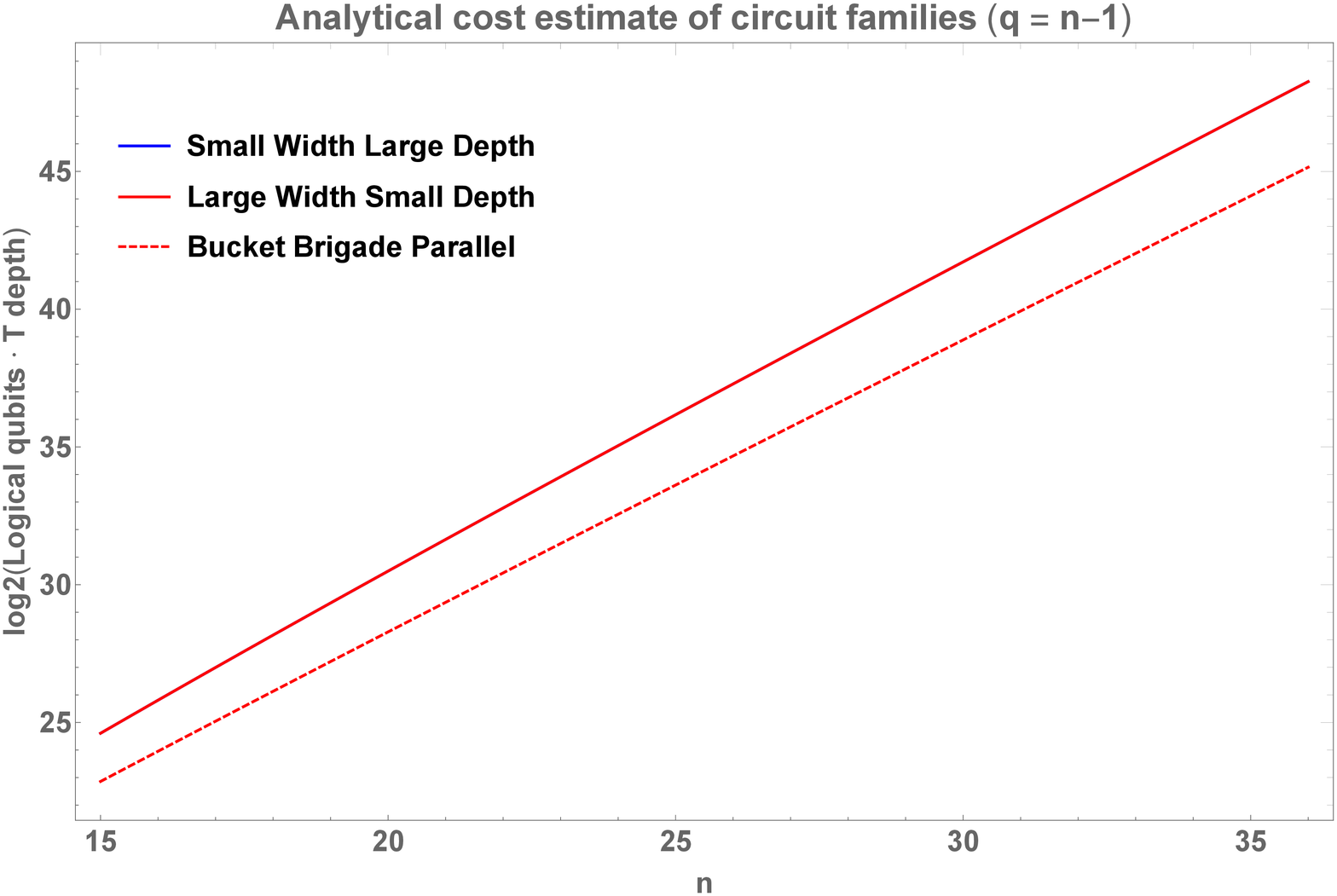}}
  \subfigure{\includegraphics[width=\columnwidth]{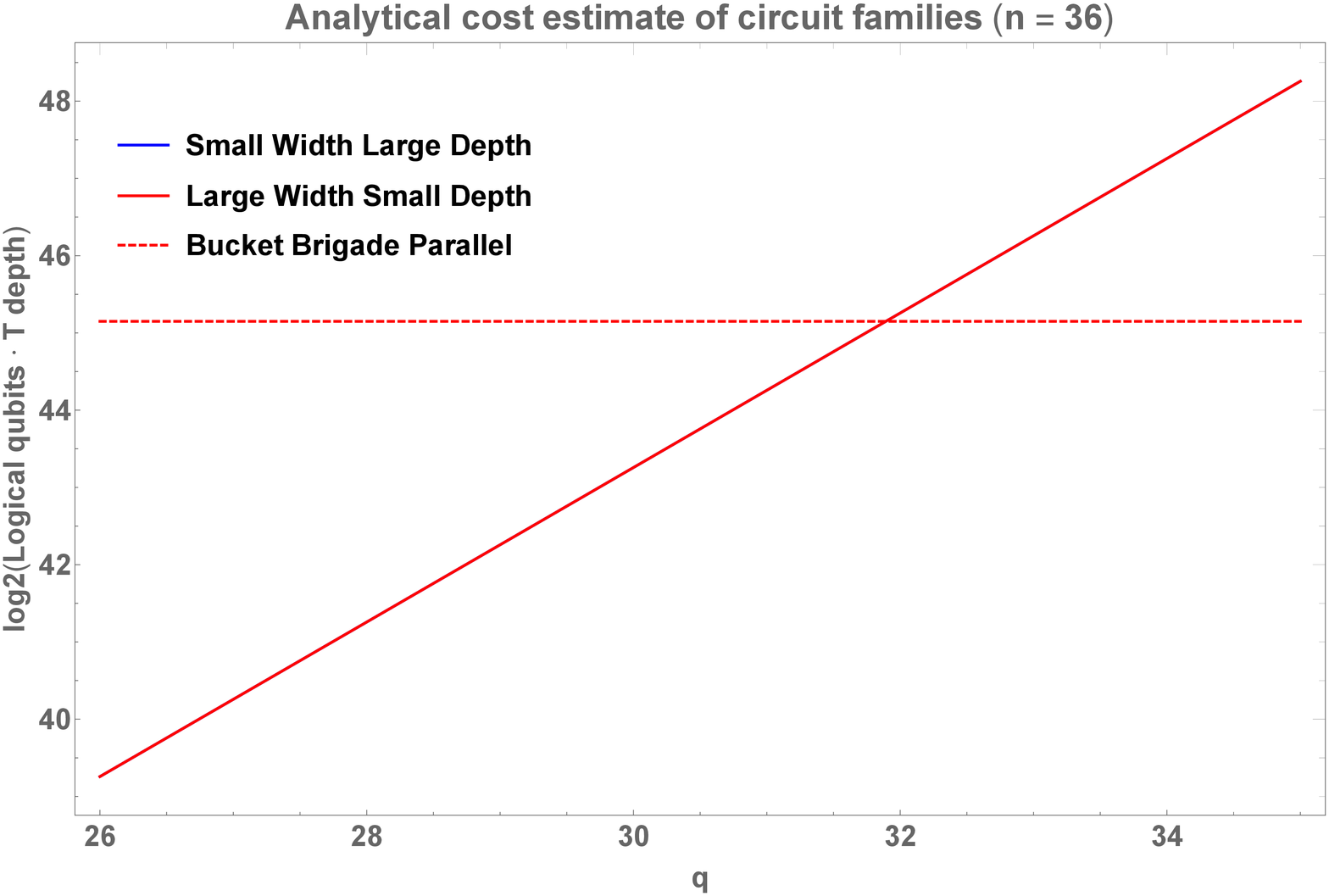}}
  \caption[Analytical cost estimates $N_Q \times T_d$ for large
  depth/width and bucket brigade circuits.]{Analytical cost estimates
    $N_Q \times T_d$ for large depth/width and parallel bucket brigade
    circuits. In both plots, the cost for small width and large width
    circuits are practically identical. (Top) Cost for $q=n-1$, a
    half-full memory. Costs for large depth/width circuits are
    comparable, showing a clear tradeoff between space and
    time. However, the parallelized bucket brigade algorithm has lower
    cost overall. (Bottom) Dependence of cost on the fullness of the
    memory (containing $2^q$ 1s). For sparser memories it may be
    cheaper, up to a point, to use a large depth/width circuit over
    the parallel bucket brigade circuit.}
  \label{fgr:analytical_costs_non_hybrid}
\end{figure}

Recall that our definition of cost is a product of space and time. We
can analyze $N_Q \times T_d$ to get a rough first estimate of how the
cost will depend on $n$ and $q$. In
\autoref{fgr:analytical_costs_non_hybrid} we plot the overall costs
(including constant prefactors) for differing values of $n$, and $q$
for the circuits in which it is relevant.  We summarize our
observations in \autoref{tab:complexity} to make it easier to see the
tradeoff between the number of qubits and the depth for each circuit.

\renewcommand{\arraystretch}{1.2}
\begin{table}[h]
  \centering
  \begin{tabular}{|c||c|c|c|}
    \hline
    Circuit & Large depth & Large width & Bucket brigade parallel \\ \hline 
    $N_Q$ & $2n$ & $n 2^{q+1} + 1$ & $ 8 \cdot 2^n $ \\ \hline
    $T_d$ & $2^{q+2}(n -2)$ & $4(n-2)$ & $2n - 1$ \\ \hline \hline
    Cost & $O(n^2 \cdot 2^q)$ & $O(n^2 \cdot 2^q)$ &  $O(n \cdot 2^{n})$ \\ \hline 
  \end{tabular}
  \caption[Cost scaling for parallel bucket brigade and large
  depth/width qRAM circuits.]{Cost scaling for bucket brigade and
    large depth/width circuits.}
  \label{tab:complexity}
\end{table}

The top panel of \autoref{fgr:analytical_costs_non_hybrid} shows the
situation of a half-full memory in which $q = n-1$. This represents a
roughly random database, and is the worst case because as mentioned
previously, if the memory is more than half-full with 1s, we can
switch the polarities of the gates to pick out the locations that are
0 instead.

For a half-full memory, the parallel bucket brigade circuit is the
best choice, as it always has lower cost. For memories that are
emptier, there is a cross-over point at
$q_{max} \approx n - \log_2 n$, before which it is in fact better to
use either the large depth or large width circuit as opposed to the
bucket brigade circuit (assuming, of course, that one has knowledge or
control over the location of the 1s in the memory).

\subsection{Surface code analysis}

We now embed these circuits into a surface code using the same
procedure as in \cite{SHAPaper}. The implementation, as well as the
data, can be found in our code repository \cite{code}. The (optimistic)
surface code parameters are input injection error probability
$p_{in} = 10^{-4}$, gate error probability of $p_g = 10^{-5}$, and a
cycle time of $t_c = 200$ns.
 
\begin{figure}[H]
  \centering
  \includegraphics[width=\columnwidth]{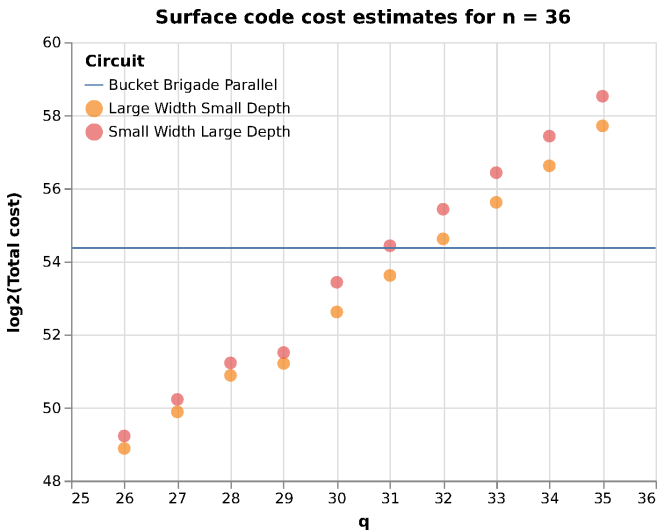}
  \caption[Cost vs. memory fullness for 8 `GB' qRAM.]{Cost vs. memory
    fullness $q$ for basic circuits with $n=36$ when embedded in the
    surface code. The horizontal line represents the parallel bucket
    brigade circuit which has only a fixed value of $n$. This matches
    closely with the analytical predictions in
    \autoref{fgr:analytical_costs_non_hybrid}, and we see that the
    cross-over point for memory fullness is around $q=31$. The bumps
    in the graph correspond to increases in surface code distances of
    the magic state distilleries.}
  \label{fig:cost_v_q_basic_bucket}
\end{figure}

\autoref{fig:cost_v_q_basic_bucket} plots the numerical equivalent of
\autoref{fgr:analytical_costs_non_hybrid}. While the relative
relationships remain the same, the overall cost is significantly
higher due to the large amount of logical qubits needed in the
distillation factories.

\begin{figure}[h!]
  \centering
  \includegraphics[width=\columnwidth]{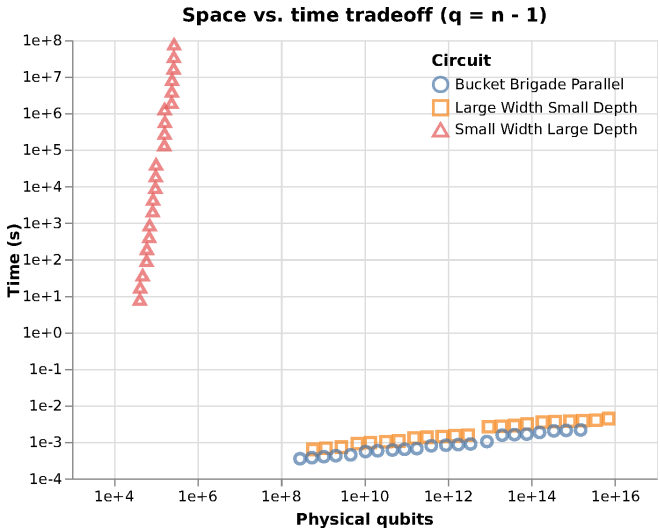}
  \caption[Space (physical qubits) vs. time tradeoff for bucket
  brigade and large depth/width circuits]{Space (physical qubits)
    vs. time tradeoff for basic circuits. Values are calculated
    assuming a surface code cycle time of 200ns, state injection error
    rate of $10^{-4}$, and intrinsic gate error rate of $10^{-5}$. We
    note that these values are quite optimistic given the current
    state of quantum hardware. Each point corresponds to a different
    memory size $2^n$ from $n=15$ to $n=36$, with smaller $n$ using
    fewer resources in each case. Memory fullness $q$ is set to $n-1$
    for each $n$ for the large width/depth circuits. }
  \label{fig:space_v_time_basic_bucket}
\end{figure}

\autoref{fig:space_v_time_basic_bucket} is perhaps the more
interesting plot, as it shows the explicit tradeoffs between the
number of physical qubits and the `real' query time. Even though we
observed on \autoref{fgr:analytical_costs_non_hybrid} that the costs
of the large depth/width circuits are comparable,
\autoref{fig:space_v_time_basic_bucket} shows us exactly how large the
space vs. time tradeoff is.
 
We present numerical data for the largest and smallest values of $n$
we chose, $n=15$ and $n=36$, in
\autoref{tab:space_v_time_basic_bucket}. The $n=36$ case corresponds
to 8 `GB' of classical data in the memory, whereas $n=15$ corresponds
to 4 `KB'. These particular choices are somewhat meaningful: 4KB was
the amount of RAM that the Apple I computer shipped with back in 1976,
while 8GB is a fairly standard amount of RAM on a laptop at the time
of writing.
 
 \begin{table}[H]
   \centering
   \begin{tabular}{|l|c|c |c|c|}
     \hline
     Circuit & $n$ & $q$ & Total time (s) & Physical qubits  \\ \hline
     Bucket brigade parallel & 15 & - & $3.48 \cdot 10^{-4} $ & $2.89 \cdot 10^8$ \\ \hline
     Large width small depth & 15 & 14 & $6.24 \cdot 10^{-4}$ & $5.84 \cdot 10^8$ \\ \hline
     Small width large depth & 15 & 14 & $7.86$ & $4.23 \cdot 10^4$  \\ \hline \hline
     Bucket brigade parallel & 36 & - & $2.13\cdot 10^{-3}$ & $1.50 \cdot 10^{15}$ \\ \hline
     Large width small depth & 36 & 35 & $4.35 \cdot 10^{-3}$ & $7.06 \cdot 10^{15}$ \\ \hline
     Small width large depth & 36 & 35 & $7.55 \cdot 10^7$ & $2.80 \cdot 10^5$ \\ \hline 
   \end{tabular}
   \caption[Space (physical qubits) vs. time tradeoff for bucket
   brigade and large depth/width qRAM circuits.]{Time and physical
     qubits required for fault-tolerant qRAM queries. The sizes $n=15$
     and $n=36$ are analogous to 4KB and 8GB memory sizes
     respectively.}
   \label{tab:space_v_time_basic_bucket}
 \end{table}

 Our analysis shows that quantumly querying the 4KB qRAM can be done
 with nearly 100 million qubits in roughly 0.35ms (with parallel
 bucket brigade), or with roughly 42000 qubits in around $8$ seconds (small
 width large depth), however the latter can only be used in cases
 where we know where the 1s in our memory are. As a reference point,
 modern-day RAMs have query times on the order of 10-20ns. To query
 this fast would first of all require significant advances in
 operational speed (recall our estimate of surface code cycle time was
 an ambitious 200ns), as well as an astronomical amount of qubits.
 
 Recall also that these numbers have been computed under the
 assumption that we have as many factories as are needed to implement a
 single layer of $T$-depth. As all the MPMCTs are the same size, and
 since they stack in the parallel version, here the $T$-width
 $T_w = T_c / T_d$ is an accurate representation of the number of $T$s
 in each layer. One could, however, adjust the number of factories
 according to available resources, and the number of physical qubits
 (time) would decrease (increase) proportionally.
 
\section{Hybrid query circuits}
\label{sec:hybrid}

We now investigate a compromise between the two circuits in
\autoref{sec:basic} by creating a sort of hybrid of the two
extremes. Our motivation is to explore a wide range of options for the
tradeoff between memory and depth to enable an algorithm designer to
choose a qRAM implementation based on available resources.

We will need a larger running example: suppose our addresses are now 5
bits, and that addresses 00000, 01001, 10010, 11011, 00100, 01101,
10110, 11111 all contain the value 1 ($2^q$ full addresses, $q = 3$).

\subsection{Circuit design}

The idea behind the hybrid circuit is, rather than checking the
validity of all $n$ bits of the address, check only the first $k$, and
then use those outputs as controls for checking the rest of the
bits. For brevity of analysis we will show here only the case where
$k < q$. Full details for the case $k \geq q$ can be found in the
accompanying code \cite{code}. We also assume $4 \leq k \leq n - 3$,
as recall the MPMCT implementations must have at least 4 control bits.

\begin{figure}[h!]
  \centering
  \includegraphics[width=0.9\columnwidth]{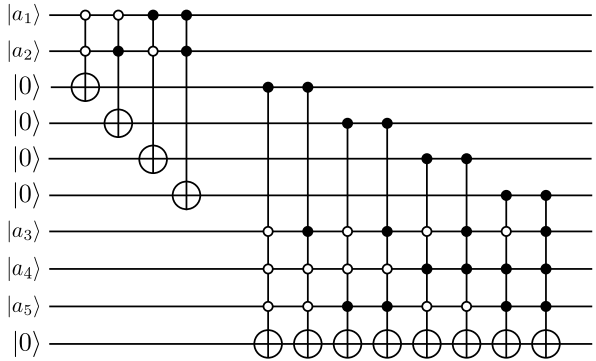}
  \caption[A hybrid qRAM circuit.]{A basic hybrid circuit. The initial
    set of controlled gates recognizes the first $k$ bits of an
    address; addresses that pass this condition go on to control
    readout of the remaining $n-k$ bits (here $n=5, q = 3, k =
    2$). Full reversibility of the query is obtained by repeating the
    first $2^k$ MPMCTs in the top tier. In general, the number of
    bottom-tier MPMCTs is not the same for each top-tier output, as is
    depicted here. In the worst case for purposes of parallelization,
    there will be $2^{q-1} + 1$ MPMCTs on one output and one on each
    of the rest, leading to more layers of CNOTs required to copy down
    the most common output.}
  \label{fgr:inbetween}
\end{figure}

The circuit for our running example is shown in
\autoref{fgr:inbetween}. In the worst case, every possible $k$-bit
string occurs as the first $k$ bits of at least one address in the
space. Thus our top `tier' consists of at most $2^k$ $k$-controlled
MPMCTs, and as a consequence, we need at most $2^k$ extra qubits to
store their results. The bottom tier will always consist of $2^q$
$n-k+1$-controlled MPMCTs, some of which will share the same control
bit from the first tier output (in the worst case, this will be
$2^{q-1} + 1$ of them, as this requires more layers of fanout when we
parallelize). Again, in specific cases we could compute the ESOP of
the Boolean expressions to simplify the products of MPMCTs, but here
we assume them to be unknown.

We continue using the decomposition of the MPMCTs in the previous
section. The number of ancillae now depends on the size of $k - 1$
vs. $n - k$ - we will need enough ancillae to implement the larger of
the two gates. As the ancillae are returned to their initial state
after use, we can use the same ancillae for all the gates in the
sequence. Finally, to make the query fully reversible, we need to run
the top `tier' a second time to return the register of $2^k$ qubits to
0.

We can use the results of the previous sections to compute
\begin{eqnarray}
 \hspace{0.5cm}
  N_Q &=& n + 2^k + 1 + \max(k - 1, n - k)  \nonumber\\
  D &=& 2\cdot2^k(28k - 60) + 2^q(28(n-k+1)-60)  \nonumber\\
  T_c &=& 2 \cdot 2^k (12k - 20) + 2^q(12(n-k+1) - 20)   \nonumber\\
  T_d &=& 2 \cdot 2^k\cdot4(k-2) + 2^q\cdot4(n-k+1-2) \hspace{0.3cm}\\
  H_c &=& 2 \cdot 2^k(4k-6) + 2^q(4(n-k+1)-6))  \nonumber\\
  \hbox{CNOT}_c &=& 2 \cdot 2^k(24k-40) + 2^q(24(n-k+1)-40)  \nonumber
\end{eqnarray}

\begin{figure}[H]
  \centering
  \includegraphics[width=0.9\columnwidth]{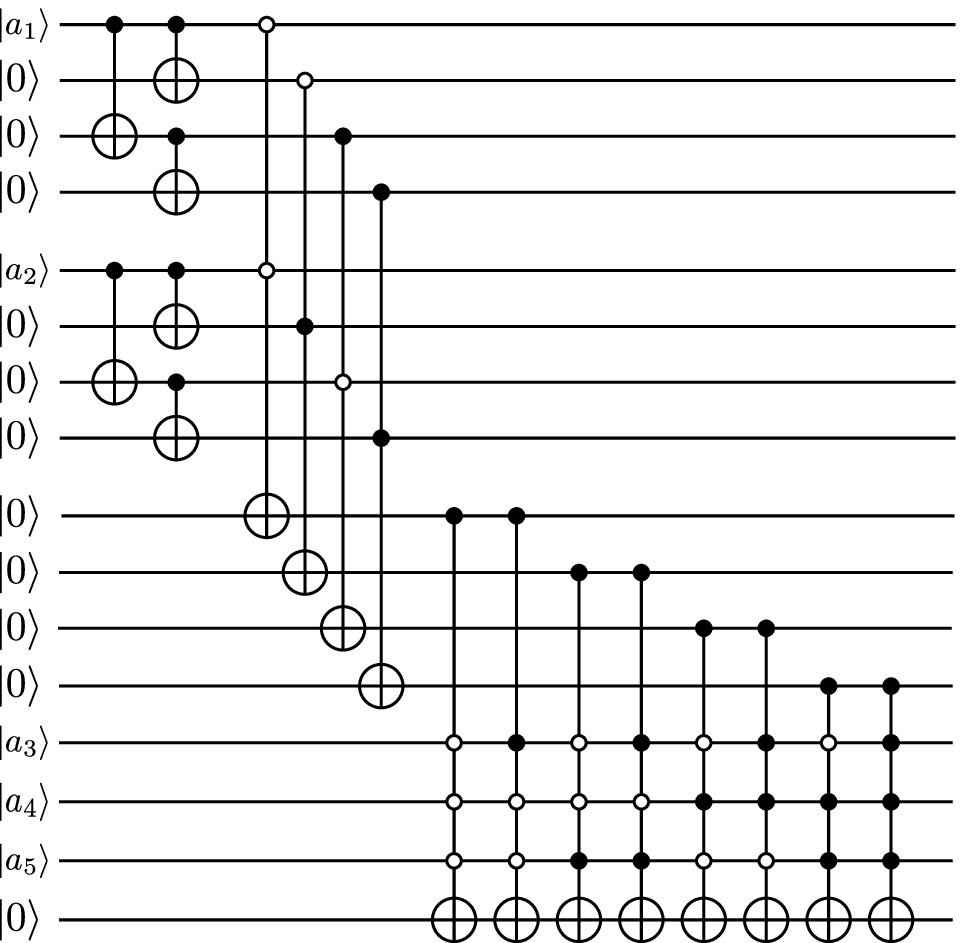}
  \caption[A partially parallelized hybrid qRAM circuit.]{A hybrid
    circuit with its first tier parallelized. Parallelizing only one
    tier of a hybrid circuit leads to an increase in cost in the worst
    case, as a larger number of qubits are required for
    parallelization, while the other tier still runs sequentially and
    may take a significant amount of time.}
  \label{fgr:hybrid_tier1_parallel}
\end{figure}

We can, in addition, parallelize the hybrid circuit in the same way
that we parallelized the original deep circuit. There are three ways
to do this: parallelize only the first tier (as shown in
\autoref{fgr:hybrid_tier1_parallel}), parallelize only the second
tier, or parallelize both tiers. Parallelizing both tiers is clearly
the best choice, as not fully parallelizing will incur additional cost
(more qubits) without seeing the full benefit in terms of time
saved. We include these other approaches only as an intermediate step
and plot them to show how they are sub-optimal. Resource counts for
the aforementioned parallelizations can be found in the code.

\subsection{Surface code analysis}

\begin{figure}[h!]
  \centering \includegraphics[width=\columnwidth]{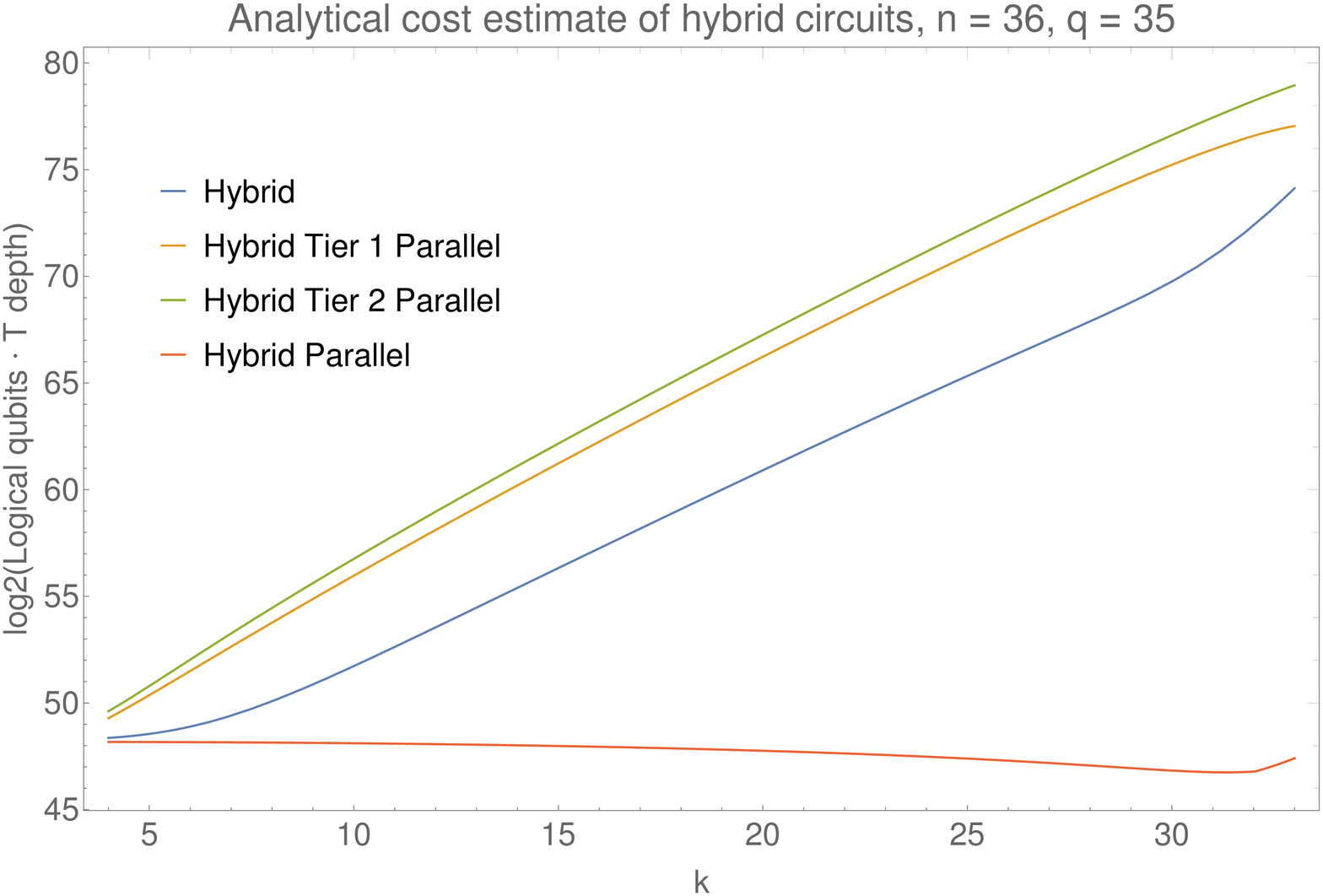}
  \caption[Analytical dependence of cost on hybrid splitting
  parameter.]{Analytical dependence of cost vs the hybrid splitting
    parameter $k$ for a memory with $n=36$ and fullness $q=35$. As
    expected, the partially parallelized versions do very poorly
    because parallelization of one half incurs a large overhead in the
    number of qubits, while running the other half still takes a
    significant amount of time. Fully parallelizing the circuit yields
    lower costs overall, and we also observe an optimal value of $k$
    for a given $n, q$.}
  \label{fig:cost_v_k_hybrid}
\end{figure}

\begin{figure}[h!]
  \centering
  \subfigure{\includegraphics[width=\columnwidth]{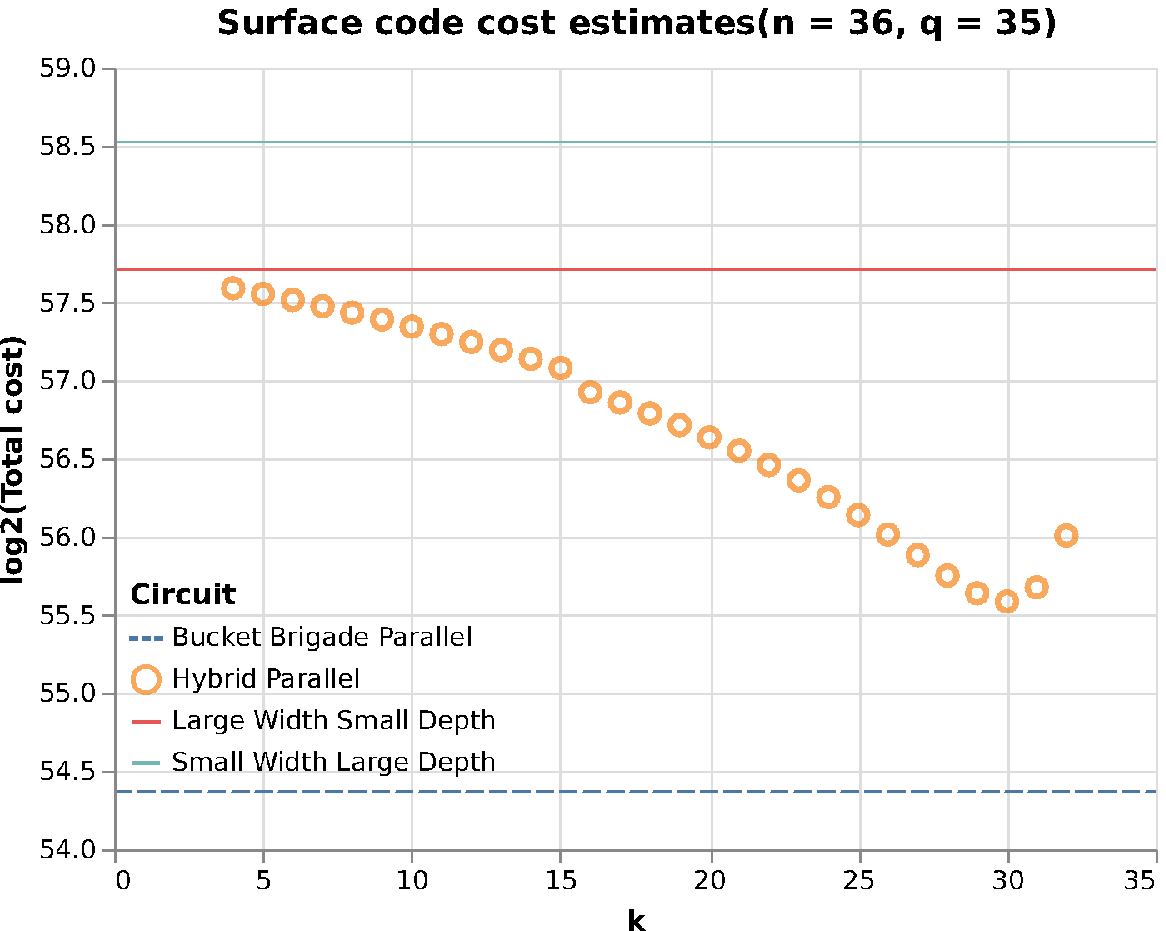}}
  \subfigure{\includegraphics[width=\columnwidth]{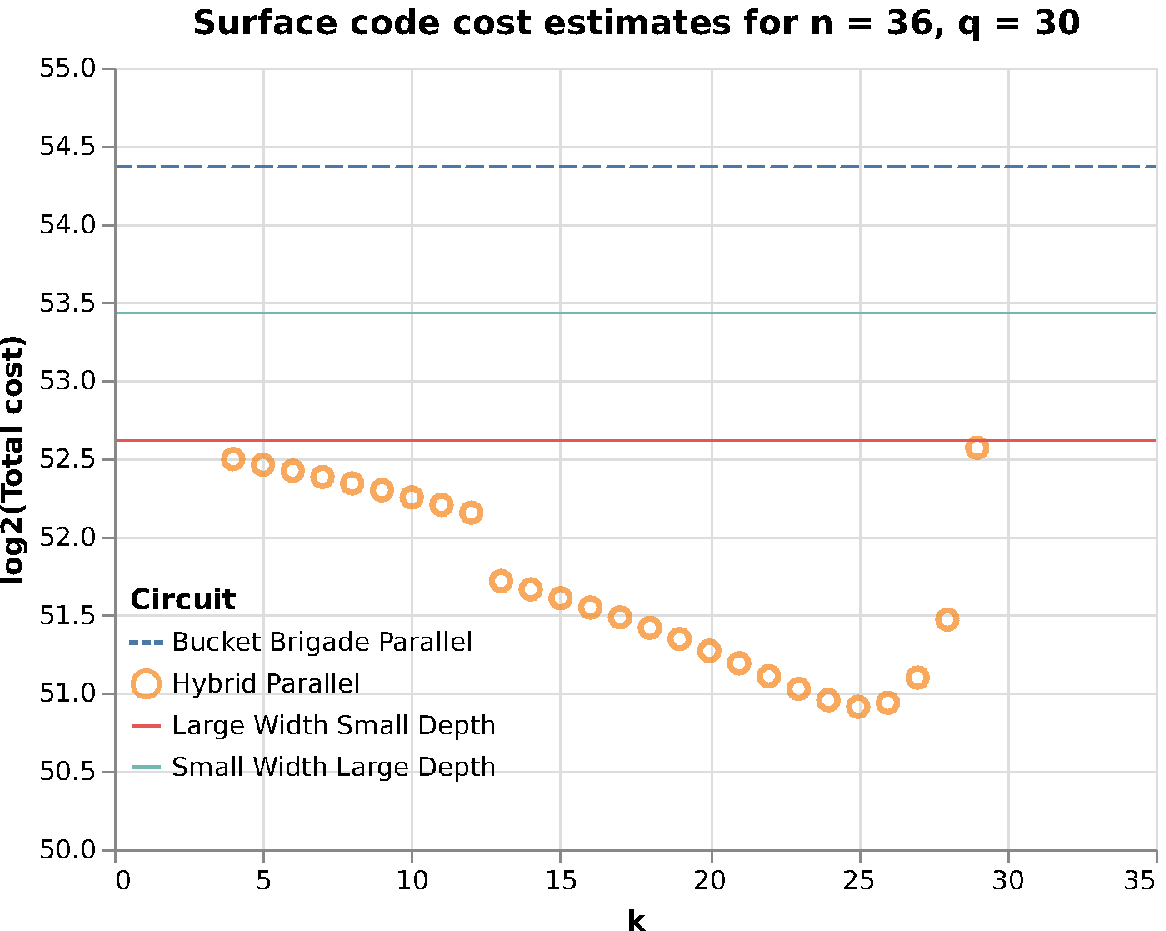}}
  \caption[Cost of hybrid qRAM circuits for different memory
  fullness.]{Surface code costs for hybrid circuits with $n=36$, and
    $q=35$, $q=30$. For fuller memories the parallel bucket brigade
    circuits are still the better choice, but for sparse memories the
    parallel hybrid circuits have lower costs than all previous
    circuit families.}
  \label{fgr:sc_hybrid}
\end{figure}

We now study the tradeoffs between $n$, $q$, and our new parameter
$k$. We will again perform this in two ways, using $N_Q \times T_d$,
as well as a full surface code analysis. The former is plotted in
\autoref{fig:cost_v_k_hybrid}, while the surface code results are
shown in \autoref{fgr:sc_hybrid} for a fixed $n$ and two different
values of $q$.

Unlike in the previous section where the dependence on $n$ and $q$
could be expressed simply as $O(n^2 2^q)$, the hybrid circuits carry a
very complex, intertwined connection between $n, q$, and $k$. In
particular, while we only ever see up to a quadratic dependence on
$n$, we see many terms with exponential dependence on $k$ and $q$,
such as $nk2^k$, $n 2^{k+q}$, etc. For the most part,
\autoref{fig:cost_v_k_hybrid} shows a clear cut exponential dependence
on $k$ when $n$ and $q$ are fixed, deviating only at the extremal
choices of $k$.

We observe that here in the worst case, when no circuit optimization
is performed, the cost of the basic hybrid circuit is actually worse
than the simple large depth/width design. Even though the size of the
MPMCTs is smaller, there are more of them, and there is the additional
exponential increase in the number of logical qubits required to store
the outputs of the first tier ($2^k$ of them). Similarly,
parallelizing only a single tier is detrimental to the overall
cost. As anticipated, the time saved in parallelizing only one part
does \ieeeedit{not} compensate for the substantial increase in the number of qubits
required to do so.

Where we do observe an improvement is with the fully parallelized
hybrid circuit. In fact the cost of this circuit actually
\emph{decreases} when $k$ is increased, and reaches a minimum that we
can solve for. For purposes of example, setting $n = 36, q = 35$, we
differentiate the product $N_Q \times T_d$ (full cost expressions can
be found in the code, and we look in the limit of $k \approx n$ based
on the graph, and due to the complexity of the expressions). Using
$(n-k)2^q$ as the maximum amount of ancillae yields $k \approx 31.4$,
which corresponds to the minimum seen in \autoref{fig:cost_v_k_hybrid}
around $k = 31$.

We attribute the decrease in cost with increasing $k$, seen in
\autoref{fig:cost_v_k_hybrid} and \autoref{fgr:sc_hybrid}, to more
efficient use of ancillae. When all the MPMCTs are performed in
parallel, we need enough ancillae to perform \ieeeedit{the larger of the two tiers, either
$(k-1)2^k$ for the first tier, or $(n-k)2^q$ for the
second}. Recall that we are considering here the case when
$k < q$. When $k$ is small and $n-k$ is large, we need a large number
of ancillae to perform the second tier, but not many of these need to
be reused when performing the first tier. On the other hand as $k$
increases and $n-k$ decreases, the number of ancillae needed for the
second tier decreases, and furthermore a greater proportion of these
can also be used for the first tier. Eventually, we see an increase
again as the number needed for the first tier surpasses that of the
second. The specific location of this increase depends on the relative
values of $n, k$, and $q$. In the case where $q = n - 1$, it can be
approximated by $k \approx n - \log n + o( \log \log n)$, which fits
with the observed minimum.

\begin{figure}[h!]
  \centering
  \subfigure{\includegraphics[width=\columnwidth]{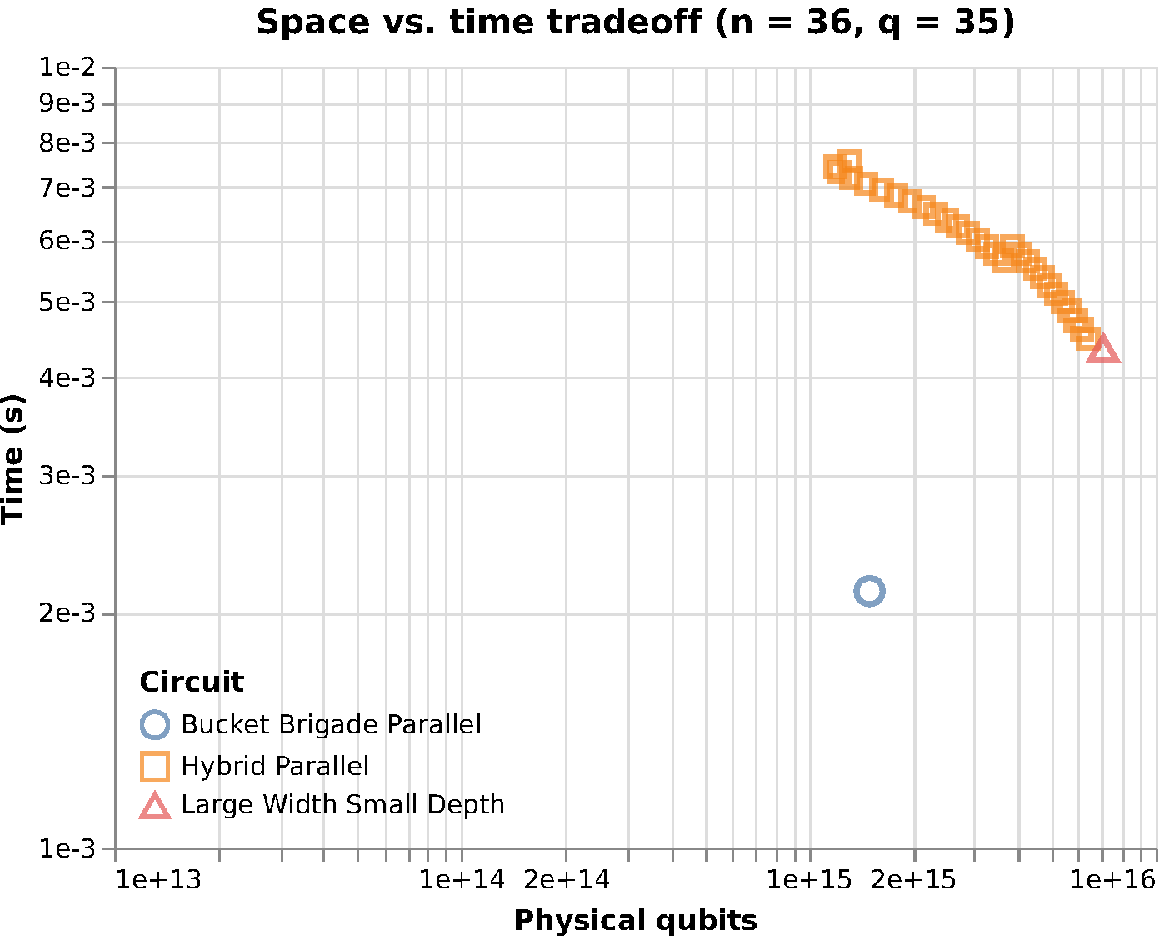}}
  \subfigure{\includegraphics[width=\columnwidth]{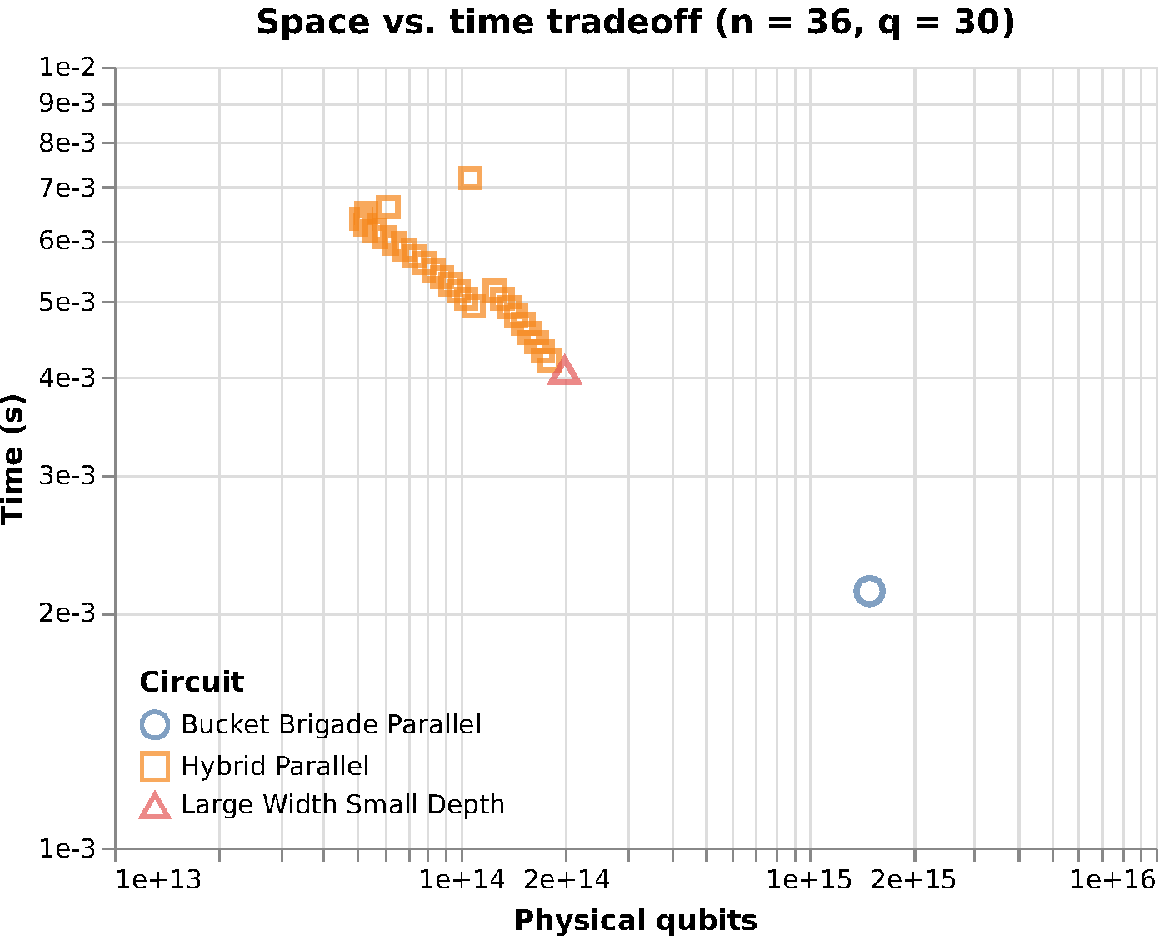}}
  \caption[Space (physical qubits) vs. time tradeoff for hybrid qRAM
  circuits.]{Time vs. physical qubits for the hybrid circuits for
    $n = 36$, and $q = 35$ ($q = 30$).  Distinct points represent
    different values of $k$ from 4 to 32 (26). Lower values of $k$ are
    at the bottom right for the hybrid parallel, and increasing first
    towards the left before reaching a minimum, and then increasing up
    to the right. This follows scaling of
    \autoref{fig:cost_v_k_hybrid}.}
  \label{fgr:time_v_physq_hybrids}
\end{figure}

\autoref{fgr:time_v_physq_hybrids} shows again the tradeoff between
physical qubits and time for the hybrid circuits. For our 8GB case, to
obtain millisecond-order query times we must still use on the order of
$10^{15}$ physical qubits in the fully parallel version, while a
million physical qubits in the basic version yields query times on the
order of 3 years.

\section{Improving bounds with optimization}

The basic circuits of \autoref{sec:basic} showcase the extremes. Without optimization, these circuits, and even the improved hybrids in
\autoref{sec:hybrid}, take a prohibitive amount of resources to execute
given the present state of hardware. Taking advantage of special
structure and optimization is therefore a critical step moving
forward, and work along these lines has already begun to be developed
\cite{Babbush2018, Childs2018, Vadym2018}. Here we analyze one such
case \cite{Vadym2018} that naturally incorporates tradeoffs in space
and time similar to the ones we have explored.

Recall that \autoref{fgr:algorithm_schematic} depicts an algorithm
that may use some of its constituent qubits to query a qRAM while the
rest remain idle. In a situation where qubits are a limited resource,
it would of course be advantageous to use these idle qubits for the
query. Such a family of circuits with this property was recently
proposed in \cite{Vadym2018}. Called \textsc{SelectSwap} circuits,
they are based on an improved method of performing a uniformly
controlled rotation \cite{Childs2018}, i.e. a sequence of
mixed-polarity gates which includes all possible configurations of the
control bits, coupled with a network of SWAP gates. While they too
depend on knowledge of the database, clearer bounds for their
execution can be derived.

\begin{figure}[h!]
  \centering
  \includegraphics[width=\columnwidth]{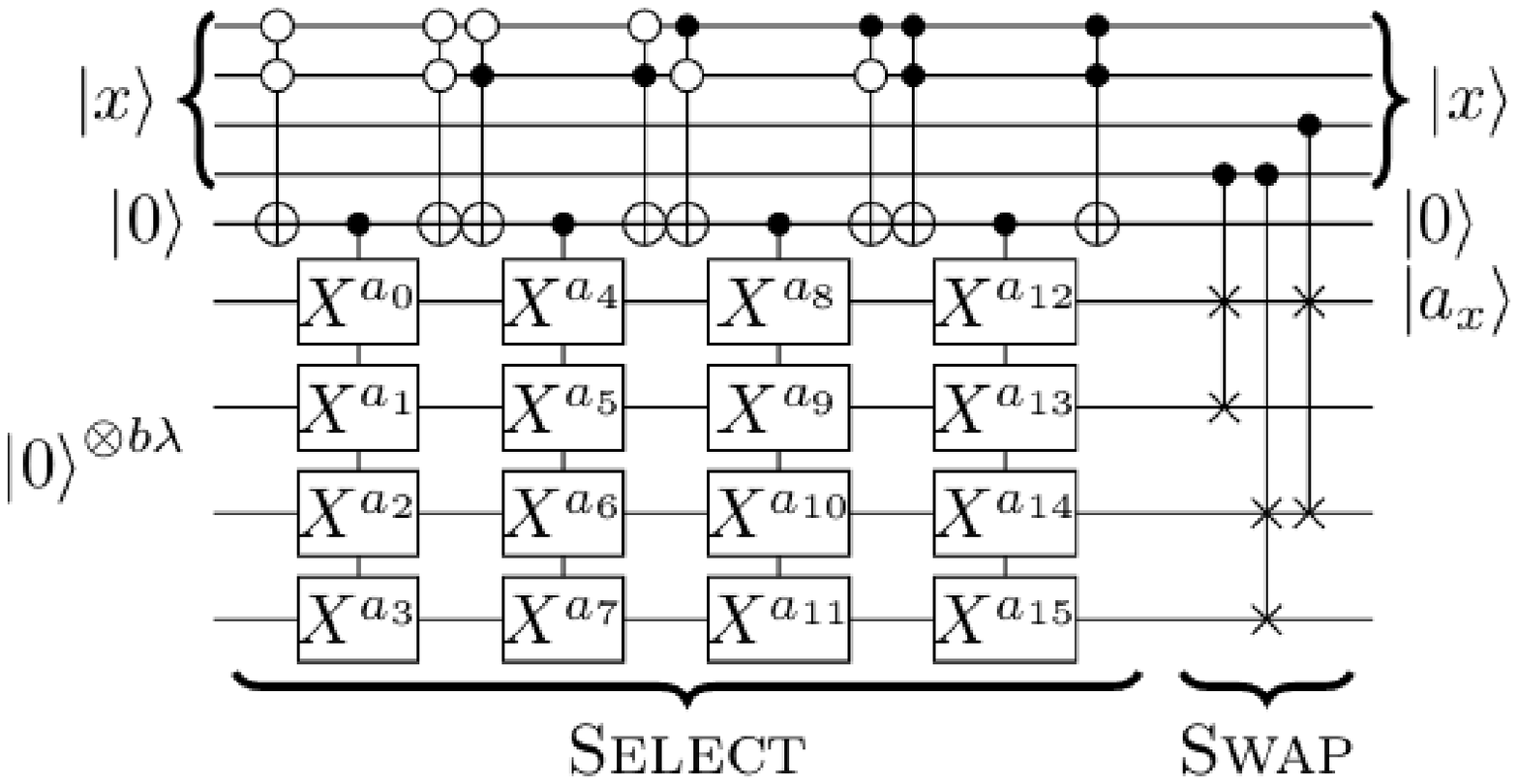}
  \caption{A \textsc{SelectSwap} circuit from \cite{Vadym2018}. The
    parameter $b$ denotes the number of bits stored at each address
    location (we take $b=1$). The values $a_i$ are either 0 or 1
    depending on the contents of the address. The \textsc{Select}
    portion of the circuit picks out the address location specified by
    $\ket{x}$, and the \textsc{Swap} portion serves to move the
    queried bits into an output register. Tradeoffs can be made by
    varying the value of $\lambda$ - increasing $\lambda$ will
    decrease the number of controls in the MPMCTs, at the cost of
    increasing the number of lower registers of qubits, which in turn
    increases the number of SWAPs.}
  \label{fgr:selectswap}
\end{figure}

An example \textsc{SelectSwap} circuit is shown in
\autoref{fgr:selectswap}, as presented in \cite{Vadym2018}. The
circuit consists of two parts, a \textsc{Select} operation where a
series of MPMCTs marks the data at the queried address, and a
\textsc{Swap} operation that moves the results to an output
register. A tradeoff is made using a parameter $\lambda$, which plays
a role analogous to the $k$ of our hybrid circuits. In a
\textsc{SelectSwap} circuit, $\lambda$ represents a number of copies
to make of an output register to receive the results of MPMCTs. A
larger $\lambda$ leads to a larger number of qubits, however it also
decreases the size and thus resource count of the MPMCTs. In addition
to the version of \autoref{fgr:selectswap}, which requires the output
registers all be initialized to $\ket{0}$, it is possible to use dirty
ancillae as the input, at the cost of needed to perform the inverse of
the \textsc{SelectSwap} operation and a sequence of additional SWAPs.

%
%

\begin{table}
  \begin{tabular}{c|c|c}
     & Clean anc. & Dirty anc. \\ \hline 
    Qubits & $b\lambda + 2 \lceil \log_2 N \rceil$ & $(b + 1) \lambda + 2 \lceil \log_2 N \rceil$  \\ \hline
    $T$-depth, $\mathcal{O}(\cdot)$ & $\frac{N}{\lambda} + \log\lambda$ & $\frac{N}{\lambda} + \log\lambda$ \\ \hline
    $T$-count, $\leq \cdot + \mathcal{O}(\log \cdot)$ & $4 \lceil \frac{N}{\lambda} \rceil + 8b\lambda$ & $8 \lceil \frac{N}{\lambda} \rceil + 32b\lambda$ \\
  \end{tabular}
  \caption{Resources required for \textsc{SelectSWAP}
    circuits with either clean or dirty ancillae. Reproduced here for convenience from Table II in
    \cite{Vadym2018}. For comparison with the circuits in previous
    sections, we take $b=1$ and $N = 2^n$ to be the size of our
    address space.}
  \label{tab:selectswap}
\end{table}

\begin{figure}[h!]
  \centering
  \includegraphics[width=\columnwidth]{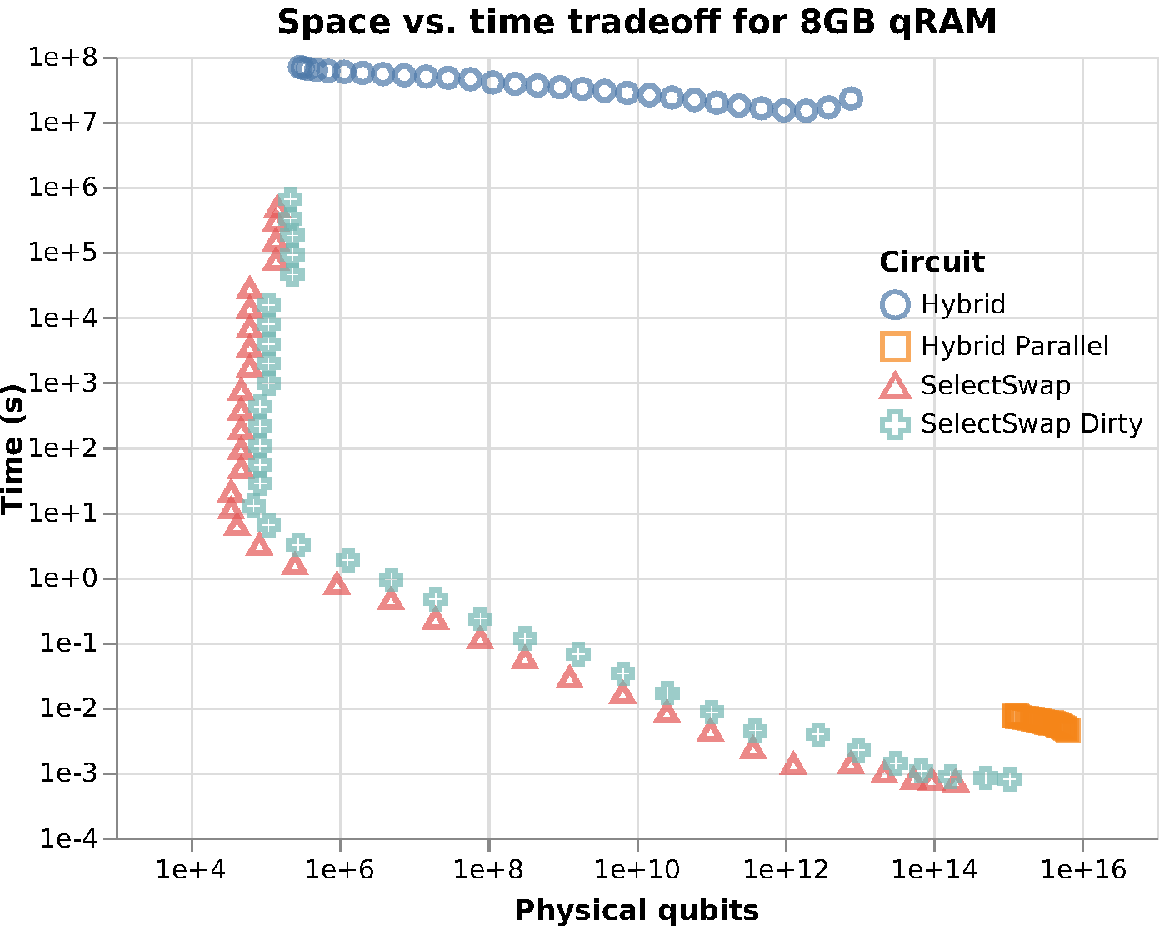}
  \caption{A \textsc{SelectSwap} circuit from \cite{Vadym2018}. The
    parameter $b$ denotes the number of bits stored at each address
    location (we take $b=1$). The values $a_i$ are either 0 or 1
    depending on the contents of the address. The \textsc{Select}
    portion of the circuit picks out the address location specified by
    $\ket{x}$, and the \textsc{Swap} portion serves to move the
    queried bits into an output register. Tradeoffs can be made by
    varying the value of $\lambda$ - increasing $\lambda$ will
    decrease the number of controls in the MPMCTs, at the cost of
    increasing the number of lower registers of qubits, which in turn
    increases the number of SWAPs.}
  \label{fgr:selswap_resources}
\end{figure}

The required resources for both versions are displayed in
\autoref{tab:selectswap}. There is a clear improvement in scaling over
the hybrid circuits, though we note that only the overall
  complexity is presented and not explicit resource counts; to make a
  proper comparison with our circuits would require knowledge of the coefficients and any higher-order terms. \autoref{fgr:selswap_resources} shows the qubit and time requirements
  for the case where $b = 1$ and $\lambda$ is chosen to be optimal at
  $\mathcal{O}(\sqrt{N/b})$. The optimal point is clearly visible. In
  the case of the dirty qubits, the execution time is slightly
  greater, however this may be a small compromise if we are able to
  make heavy use of dirty qubits in the rest of the algorithm.

While such analysis shows improvement at the logical level, it is not
clear such savings will fully translate to the fault-tolerant
level. For example, if the query circuit and algorithm circuit are
embedded in different error correcting codes, it might not be possible, or worthwhile 
to translate between the two in order to use the algorithms dirty qubits.


\section{Conclusion}

We have presented a number of different circuit families that perform
the task of a qRAM. It is important to note that our resource
estimates are based on the worst-case situations of each. One should
always, of course, do what's best for the problem at hand. For a
specific algorithm, application of circuit synthesis and optimization
techniques may yield lower cost for, e.g. one of the
partially-parallelized hybrid circuits rather than the
fully-parallelized versions.

Regardless, we can still draw some interesting conclusions from our
analysis. First, unsurprisingly, to implement a fault-tolerant qRAM
with as much logical memory as a current-generation laptop is
infeasible under our physical assumptions and the current state of
quantum hardware and quantum fault-tolerant error correction. Under
our (currently optimistic) assumptions and using current
fault-tolerant quantum error correction, an 8GB qRAM that is roughly
half-full uses quadrillions of qubits to obtain fast query times on
the order of milliseconds; alternatively, working with only millions
of physical qubits would yield query times on the order of
years. While circuit optimization may alleviate this to some degree,
except for cases with non-trivial structure, we do not anticipate
this will shave off enough orders of magnitude to make a
fault-tolerant circuit-based qRAM of this size feasible in the
foreseeable future.

We also note that while our main analysis assumes generic unstructured
data, substantial optimizations are possible when there is special
structure in the data. For example, if bit
$b_{i}, i \in \{0,1, \ldots, 2^n -1\}$, is known to be a function of
bits $c_{j}, j \in \{0,1, \ldots, 2^m - 1 \}$, where $m << n$, then
depending on the complexity of computing $b_i$ from the relevant
values of $c_j$, it may be most efficient to build a qRAM for the $c$
values and compute the $b$ values in the circuit making the queries. A
particular example, discussed in \cite{DiMatteo2018} and included in
our software, is the case where the address space has Cartesian
product structure. We can take advantage of this to write a circuit
that queries two smaller qRAMs and then check the validity of both
parts using a Toffoli gate. Finding such structures that are useful in
practical is an interesting area for future work.

One significant opportunity for improvement is in the implementation
of the surface code. Lattice surgery has recently been shown to yield
a decrease in resource estimates, in some cases lowering the number of
physical qubits by a factor of 4 to 5~\cite{Horsman2011, Litinski2018,
  Fowler2018}. While a factor of 5 may not have much of an effect on
circuits requiring quadrillions of qubits, it is promising for smaller
qRAMs requiring on the order of 10000 qubits. Further improvements in
fault-tolerant methods as well as advances in experimental techniques
for reducing physical error rates may make small qRAMs feasible in the
nearer term.

\section*{Acknowledgments}

We thank Matthew Amy, Austin Fowler, Craig Gidney, Alexandru Paler, and Mathias Soeken for useful discussions. We also thank
Dominic Berry and Craig Gidney for directing us to reference \cite{Vadym2018}.  We
acknowledge support from NSERC and CIFAR. IQC and the Perimeter
Institute are supported in part by the Government of Canada and the
Province of Ontario.

\bibliographystyle{unsrt} \bibliography{tempbib}

\end{document}